# Building Trust: Foundations of Security, Safety and Transparency in AI

Huzaifa Sidhpurwala
huzaifas@redhat.com

Garth Mollett
gmollett@redhat.com

Emily Fox
efox@redhat.com

Mark Bestavros
mbestavr@redhat.com

Huamin Chen
hchen@redhat.com

## Abstract

This paper explores the rapidly evolving ecosystem of publicly available AI models, and their potential implications on the security and safety landscape. As AI models become increasingly prevalent, understanding their potential risks and vulnerabilities is crucial. We review the current security and safety scenarios while highlighting challenges such as tracking issues, remediation, and the apparent absence of AI model lifecycle and ownership processes. Comprehensive strategies to enhance security and safety for both model developers and end-users are proposed. This paper aims to provide some of the foundational pieces for more standardized security, safety, and transparency in the development and operation of AI models and the larger open ecosystems and communities forming around them.

# Introduction

## Artificial intelligence and Generative AI

Generative AI, a branch of artificial intelligence focused on AI production of content such as text, images and video, has seen significant advancements since the introduction of generative adversarial networks (GANs) in 2014 (Goodfellow et al., 2014), which improved data generation but faced issues like training instability. The development of transformers and self attention mechanisms in 2017 (Vaswani et al., 2017) facilitated further improvements in natural language processing, leading to large language models (LLMs) like GPT (Radford et al., 2018) with highly advanced text generation capabilities. Diffusion models (Sohl-Dickstein et al., 2015) have also seen rapid advancement in image and video generation.

This rapid advancement in technology capability has been matched by an equally rapid uptake in adoption. Forbes predicts the AI market will reach a staggering $407 billion by 2027[1],. Additionally, organizations looking to tap into this high-growth market are driving rapid development and innovation in this space to achieve competitive advantages, streamline operations, and enhance customer engagement in an increasingly digital landscape, particularly focusing on the development of models and model systems that address emerging use cases in sectors like manufacturing, service operations, and marketing and sales[2], promising cost savings as a key benefit.

---

[1] https://www.forbes.com/advisor/business/ai-statistics/
[2] https://indatalabs.com/blog/ai-cost-reduction

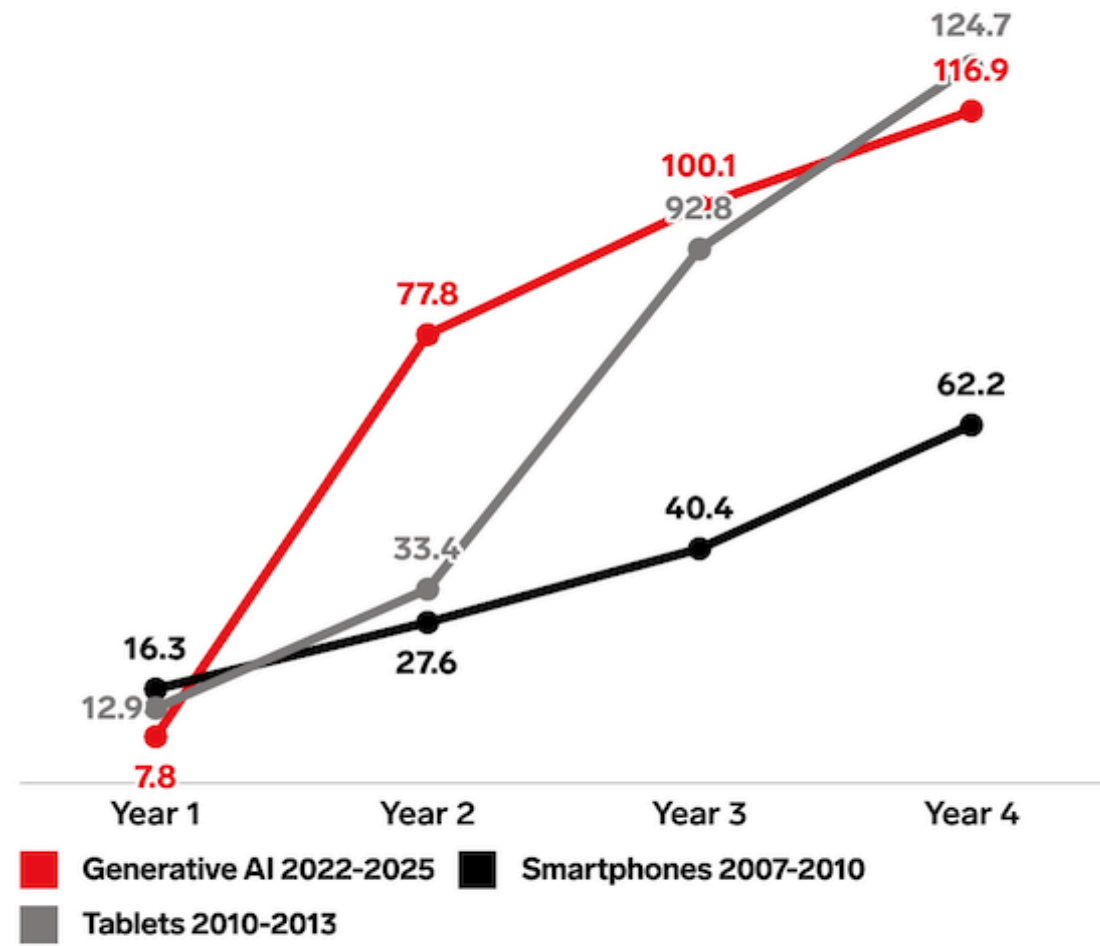

Image courtesy of https://www.emarketer.com

As with any new technology, it is worth noting that the industry is still identifying new and valuable uses for AI and these market predictions may fluctuate as use cases are tested in real world environments with real world problems.  Other factors such as the market demand for GPUs outpacing the current supply and potential market control concerns by regulators, may impact these projections.

There are various terminologies and efforts[3] in the industry to define open source models.[4] However a discussion of these efforts and research[5] in this field is beyond the scope of this paper. For the purpose of clarity we shall be using the term **public model**, *for a model which is publicly available for download and use.*

## LLMs: data versus code

LLMs are the next evolution of data science, a field focused on math and data. Unlike traditional systems and applications which rely on logic and programming for a specified outcome, large language model development typically consists of architecture research and design, which is then coded. In most cases, PyTorch or any of the other deep learning libraries are used, followed by training in multiple stages. Post-training, model weights are saved in a file that contains billions of, most commonly, 16-bit floating point numbers (or low precision data times to reduce computational and memory costs by using quantized models). These weights are then shipped to consumers of LLMs, along with configuration files and other metadata. Therefore in their very simplest form, these AI models are mostly data with some code as compared to traditional systems, applications, packages and libraries, such as Python or binary executable files.

Popular model hosting services, like Hugging Face[6] host models in the "safetensors" format, allowing parameters to be stored for safe load[7] and operate.[8]

When the model is used for inference, such as generating output, these weights are loaded by previously used deep learning libraries; PyTorch is the example in this case. This process is also referred to as "operationalizing a machine learning model" or "putting a machine learning model into production."

LLMs present a new combination of considerations, risks, and concerns not seen in industry to date. In addition to the application of existing security constructs (vulnerabilities and controls) the data and intended use of LLMs bring safety concepts historically reserved for content management and social media platforms to the forefront of risk discussions. This distinction in both their development and operation, as compared to traditional systems and applications, imply special attention and evaluation must be taken in proposing an industry wide approach to the security and safety of LLMs.

---

[3] https://www.redhat.com/en/topics/ai/open-source-llm
[4] https://opensource.org/deepdive/drafts/open-source-ai-definition-draft-v-0-0-9
[5] https://www.ibm.com/think/topics/open-source-llms
[6] https://huggingface.co/models
[7] https://huggingface.co/docs/transformers/en/main_classes/model
[8] https://github.com/huggingface/transformers/tree/main/src/transformers/models

## How AI Security is different from AI Safety

AI Security and AI Safety are interrelated yet distinct aspects of managing risks in artificial intelligence systems. AI Security primarily focuses on protecting AI systems from technical threats (both internal as well as external), unauthorized access, and attacks, which traditional information technology security typically covers. For models, this encompasses safeguarding data integrity, ensuring confidentiality of the data generated by these models, and maintaining the availability of AI services and environments for training and tuning. It involves implementing robust security measures to prevent adversarial attacks, data breaches, and other forms of exploitation that could compromise the functionality or integrity of the AI models and the underlying infrastructure.

AI Safety is typically concerned with ensuring that AI systems, LLMs, and the supporting libraries and software like PyTorch and its inference software, and their data lifecycle practices ensure the intended operation, as defined, does not cause unintentional harm to users, society, or the environment[9]. Alignment with human-defined values and accepted outputs to mitigate this harm involves addressing issues such as algorithmic biases, ensuring reliable and predictable behavior, and designing fail-safes to mitigate potential harm, negative consequence, and any potential risks presented. While AI Security protects the system from external and insider threats, AI Safety ensures that the system and the data do not pose a threat or create harm through its operation due to its development, training, and use (intended or not).

In the realm of generative AI models, the line between security and safety issues often blurs significantly. For instance, a prompt injection attack, typically considered a security concern, can lead to safety issues such as the model producing toxic or harmful content, or exposing personal information such as PII and PHI. This intersection highlights the critical need for a comprehensive approach to AI risk management that addresses both security and safety concerns in tandem.

---

[9] Sustainable AI is often considered its own independent domain as the impacts of AI on the environment are prolonged and often unable to forecast due to additional factors. It's reasonable to state, however, that the resulting impact is cumulative given current studies for traditional CPU usage. Therefore the authors consider Sustainable AI as a subset of AI Safety in that the ecological and corresponding human harm that results from unsustainable AI development and usage substantially influences the duration of its benefits for humans against environmental realities of changing climate.

Both aspects are crucial for the responsible development and deployment of AI technologies, and their interconnected nature underscores the importance of a holistic approach to AI risk mitigation.

## Effectively navigating AI Security and safety issues

It is essential that organizations developing or using AI establish appropriate processes and workflows that address the distinct aspects and considerations necessary for AI Security and for AI Safety within an overall AI Risk Management function. While there are areas of reuse and adaptation that are relative between these domains, the overall intent and outcome of effectively managing AI Security and AI Safety is to reduce the risk and subsequent impact of both on an organization, users, and society at large. Therefore, it is recommended that distinct yet parallel structures be established that allow for concentrated management of issues in each domain enabling areas of collaboration and corroboration between those areas.

In the context of AI Safety and AI Security, this adds another dimension to the fact that secure models may facilitate more safe models (through transparency, metadata, and other security requirements on integrity and confidentiality), just as Safe models, by nature of their practice, could be secure (applying best practices of safety - output verification, etc.) the probability of a safe and responsible model not having some semblance of security is low given the focus of "doing the right thing" which is instrumental to achieving safety outcomes.

### Objectives, methodologies, and interdisciplinary expertise

AI Security focuses on protecting AI systems from external threats to confidentiality, integrity, and availability, while AI Safety deals with assuring AI behaves as intended and does not cause harm. These different objectives may require distinct methods for discovery, triage, remediation, and management reflective of their core focuses, desired outcomes, and inherent expertise required to execute and reason about those issues; suspected and reported.

However, it may occur that the exploitation of a security vulnerability allows for the model's alignment and value learning to be compromised, thereby resulting in real safety hazards. This fundamental difference in approach but interconnected impact necessitates separate considerations and specialized research efforts, or perhaps as the authors suggest, exploring shared regimes and research which favor both security and safety for well orchestrated and holistically-managed risk.

As with objectives and methodologies, AI Security and AI Safety require different sets of technical skills and interdisciplinary collaborations. The "Concrete Problems in AI Safety" paper highlights that AI Safety issues often involve complex problems at the intersection of computer science, philosophy, and cognitive science (Amodei, D., Olah, C., Steinhardt, J., Christiano, P., Schulman, J., & Mané, D. (2016).

These skills historically have not been present within the Security community, but do exist in trust and privacy teams for content platforms and social media organizations. In contrast, AI Security issues typically span multiple domains requiring specialized expertise in adversarial machine learning and secure model development as well as more traditional security concepts like defensive cryptography, network security, identity and access management, and vulnerability management. This difference in technical complexity and interdisciplinary requirements underscores the need to ensure the correct expertise is available and applied to risk mitigation and remediation for each under their respective focus that may be exchanged and re-leveraged as part of an AI risk management program.

## Temporal considerations

AI Security vulnerabilities can be considered to be static because they are already present and simply require discovery and exploitation to be actualized. They exist at the time they are coded into the software or model, and only require exploitation in the context of the operational or runtime environment for the threat to be realized, requiring immediate and rapid response upon discovery and reporting to reduce impact and mitigate damage. In effect, this is no different than security vulnerabilities in software.

Safety hazards are dynamic, in that, as the model operates, its output may stabilize or fluctuate against societal considerations and evolution. As such, they are measured on a spectrum that may evolve. Safety hazards may be realized at time of release or years later – mimicking the experience of a logic bomb, with the core difference that they are not coded in, rather derived from their training and inference as expectations and understanding of safety evolve. Therefore, safety hazards tend to be more long-term in nature.

Just as the facets of a vulnerability (i.e. maturity, impact, complexity, etc.) determine where the measure of criticality falls on a scale, most safety hazards are measured on a similar yet bracketed spectrum which requires understanding not only of the hazard, but of any shift in human-value alignment and societal expectations since trained or inferred at the time of use or harm. Another important dimension is the multilingual abilities of an LLM and the challenge of safety around this.[10]

---

[10] https://aclanthology.org/2024.findings-acl.349.pdf

Adding complexity to safety hazard triaging efforts, some generated output may be offensive to some groups of a given society while not for other groups within the same or in a different society or at a different time (as society and culture evolves in their understanding and tolerances).

These temporal considerations are one of the more distinguishing traits between AI Security and AI Safety underscoring the pressing need for safety hazard specific management, triage, and resolution processes and workflows. These should be adapted and modified from well-known vulnerability management processes and techniques for inclusion in a comprehensive AI Risk management program that encompasses both AI Security and AI Safety.

## Stakeholder involvement

AI Security vulnerabilities typically involve a narrower range of stakeholders, primarily focused on cybersecurity experts, AI developers, and organizations deploying AI systems. Comparatively, safety hazards require broader societal involvement, including ethicists, policymakers, and the general public. This difference in stakeholder composition and engagement processes supports the need for distinct resolution and management of safety hazards within a comprehensive risk management program.

## Regulatory and governance frameworks

AI Security issues often fall under existing cybersecurity regulations and compliance frameworks, while AI Safety issues will require new or adapted governance structures. While both regulatory and governance landscapes establish a need for different policy approaches, such focused processes must roll up under AI Risk management for organizations to effectively manage comprehensive risk introduced to their business or organization as a result of AI use.

Managing AI Security and AI Safety separately will result in unidentified risk actualization, ineffective coordination, and negatively impact organizations across their bottom line, brand, and market space, eroding or demolishing trust in one way or another, regardless of the innovation employed in constructing such governance. Frameworks such as NIST's AI Risk Management Framework[11] (RMF) are a good start but lack sufficient detail in pulling AI Security and AI Safety processes and workflows together to demonstrate how each influences the overall risk AI presents. Here, the application of existing risk management frameworks fail to consider the complexity of unique challenges AI presents. While NIST's AI RMF includes multiple facets of consideration, such as intellectual property which do present risk but solely in a business context, they do not necessarily consider the externality of that risk in operation as occurs with AI Security and AI Safety. Regulations such as the EU AI Act provide more clear outcomes expected of AI risk management driven by categorizing some model use as High Risk but do not provide direction on how those outcomes may be achieved. More importantly, the premise of the act defines High Risk Models as increased risk of "harm to the health, safety or fundamental rights of natural persons" fully focused on AI Safety only. This not only creates a greater rift in effectively managing AI Security and AI Safety processes to reduce or mitigate risk, it allows for considerable variability in implementation and difficulty in enforcement similar to that which was experienced with GDPR.  It is almost certain that there will be considerable effort and time spent in interpreting the Act into engineering features, checks, and tooling for models in order for enforcement to be feasible, and even then will be limited to coverage of AI Safety only.

---
[11] https://nvlpubs.nist.gov/nistpubs/ai/NIST.AI.600-1.pdf

## Current industry trend on handling security and safety issues

While the AI industry has taken steps to address security and safety issues, several key challenges still need to be addressed. Emerging trends suggest that current efforts may need to target these issues' underlying causes fully. The following examples highlight areas where further attention is needed to ensure the effectiveness of AI Security and safety initiatives.

**Prioritizing Speed Over Safety:** In the race to develop and deploy AI technologies for market share, many companies are prioritizing speed to market over thorough safety testing and ethical considerations, particularly as frameworks to address these are still under rapid development with a few in proof-of-concept or introduced by other entities[12] but not widely adopted. This is true of any early, disruptive technology, and AI is no different[13]. There is generally wide public acceptance of risks associated with technology that are typically considered acceptable trade-offs for our dependence upon them. However in the matter of Artificial Intelligence, globalization and accessibility of technology, and the impact of digital content and media on daily life, we collectively lack the years of experience in this nascent technology to create reputable AI for widespread public and business acceptance of the risks, which as of yet are immeasurable to human safety. As seen with past security movements[14], security is often years behind any technology, typically requiring a major incident before the industry begins self-correction. In the matters of AI, it is reasonable to predict that in the absence of individuals pushing for governance and risk management, we may yet experience a safety *and* security critical incident. While new models are being introduced that do prioritize safety over speed to market, lack of consensus on how to convey necessary Safety transparency information makes these challenging to comparatively evaluate. Regardless, the increase in availability of these Safety conscious models are steps in the right direction.

---

[12] https://mlcommons.org/benchmarks/ai-safety/ ,
https://www.congress.gov/bill/117th-congress/house-bill/6580/text , https://www.edsafeai.org/safe ,

[13] Reference: Wynne, Brian. (1983). Redefining the issues of Risk and public acceptance, the social viability of technology. Butterworth & Co.

[14] https://www.darkreading.com/vulnerabilities-threats/adapting-post-solarwinds-era-supply-chain-security-2024

**Inadequate Governance and Self-Regulation:** The AI industry has largely relied on voluntary self-regulation and non-binding ethical guidelines, which have proven insufficient in addressing serious security and safety concerns. This is not exclusive to AI; looking at the historical shortcomings of technology self-regulation, many governments are proposing legislation to increase the overall security of technology as can be seen with the EU's Cyber Resilience Act (CRA) among others. For AI, we see legislation proposed that doesn't align with the realities of the technology industry[15] or that addresses concerns raised by industry groups to provide meaningful outcomes[16]. Many corporate AI ethics initiatives lack teeth and fail to address structural issues or provide meaningful accountability, being developed bespoke to the entity enacting them. The lack of industry consensus – as evidenced by over 10[17] different entities focused on related and overlapping areas of AI Security and AI Safety alone – further prevents regulations from being developed that provides AI developers successful implementation and consumer enforcement. The recent veto of SB 1047[18] highlights some, but not all of these challenges.

Relatively successful self-governance within the tech industry does happen occasionally, and may also be helpful to study. Two notable examples within the security space are the gradual, widespread adoption of HTTPS by default[19] and the rapid adoption of signing technologies in the wake of the 2020 SolarWinds software supply chain compromise[20]. While both phenomena were driven by various factors, a strong common link is the availability of easy-to-use tooling: with HTTPS, easily-attainable, free certificates from e.g. LetsEncrypt helped drive adoption; with software supply chain, availability of simple signing tools such as Sigstore helped signing become a simple added step instead of a laborious additional process. For the purposes of the topic at hand, as tooling for AI Safety matures and becomes more well-known, it's possible that widespread adoption will promote greater safety as time goes on.

---

[15] https://www.gov.ca.gov/2024/09/29/governor-newsom-announces-new-initiatives-to-advance-safe-and-responsible-ai-protect-californians/
[16] https://www.eff.org/deeplinks/2024/08/no-fakes-dream-lawyers-nightmare-everyone-else
[17] OpenSSF, AI Alliance, CNCF, LF AI & Data, CISA, NIST, MITRE, OWASP, CSA, and many others
[18] https://www.gov.ca.gov/wp-content/uploads/2024/09/SB-1047-Veto-Message.pdf
[19] HTTPS encryption on the web: https://transparencyreport.google.com/https/overview
[20] Sigstore Graduates: A Monumental Step Towards Secure Software Supply Chains. https://openssf.org/blog/2024/03/20/sigstore-graduates-a-monumental-step-towards-secure-software-supply-chains/

More generally, successful industry-driven software security initiatives such as those mentioned earlier tend to involve a defined set of best practices that can be implemented relatively independently of primary feature development. This may prove difficult for the AI space: often, advancing the capability of a model may directly impact its safety, and conversely, ensuring model safety may involve holding back capability. With the industry focused on rapid advancement, prioritizing safety and security at the expense of capability may be a tradeoff key stakeholders will be unwilling to make.

**Inadequate processes for handling flaw and hazard reports, silent fixes, discouraging reporters:** There exists a lack of common methods and processes of handling model flaws reported by users - no such methodology has been proposed or shared among model makers and developers to serve as a common reporting framework outside of HuggingFace's "discussions" feature or "report model" flag which lacks sufficient documentation to independently inspect the reporting process and reasoning for reporting[21]. It has taken the software industry decades to develop a flawed-yet-functional disclosure and reporting system for software vulnerabilities, of which many model makers have little to no exposure to or experience with. AI is a technical evolution of data science and machine learning, principally distinct from traditional software engineering and technology development due to its focus on data and math and less on building comprehensive systems for users which has established methodologies for threat modeling, user interaction, and system security. Most model makers who provide their models for free may have their own bespoke processes which are obscured and may be inadequate, or may not have any process at all. In those cases, reporting an issue may involve directly reaching out to the model maker, which can be cumbersome. It is important that reporting mechanisms align to Industry's existing expectations in reporting and management of flaws — regardless if they are unique to security or safety as both require the context of the issue's discovery, disclosure of the steps by which it was found for reproducibility, evidence or indications of impact in order to simply verify the validity of the issue beyond the individual's claim in reporting. Further, both require time to reproduce (dependent on the mutations of outputs in use), develop mitigation or fixes for, coordination in release, tracking to closure, and potential development of regression testing to prevent their further recurrence.

Based on the above, we propose processes for handling both security issues as well as safety hazards for public models in the following parts of the paper.

---

[21] https://huggingface.co/docs/hub/en/moderation#reporting-a-repository does provide some information under the Moderation heading of Hub, and launches a public discussion on the topic automatically.

# Adapting existing processes to AI Models

## Traditional application security process in brief

The Software Development Lifecycle process traditionally involves designing, writing the actual application (using some type of programming or scripting language), compiling it if required, testing, and then finally packaging and shipping the end product to its users.

The development processes differ significantly between open source and closed source software. Closed source applications operate under specific expectations regarding usage and security, dictated by the organization behind them. These expectations often stem from establishing a brand and meeting customer demands for security, reliability, and support. In contrast, open source projects tend to emphasize functionality and the security boundaries of the application, relying on a collaborative model where transparency and community involvement shape the development practices. This difference means that open source applications may not face the same pressure to conform to external brand expectations, allowing for a focus on innovative features and flexibility in meeting user needs.

Despite these distinctions, there are universal industry expectations for the management of security vulnerabilities that apply to both open and closed source applications. Regardless of the development practices or the perceived robustness of the security measures in place, all software must address vulnerabilities comprehensively. The critical nature of security in software development means that organizations can face scrutiny and repercussions for failing to manage vulnerabilities effectively. Therefore, understanding and adhering to these industry standards is crucial, as users and stakeholders expect a baseline level of security from all software, irrespective of its source.

Effective vulnerability management is essential and must be implemented using industry-accepted processes, regardless of whether the software is open or closed source.

Each application security flaw is assigned a unique number by a CVE (Common Vulnerabilities and Exposures) naming authority (CNA) governed by a central, neutral and non-government affiliated organization, often considered a 'public good'. This organization's CVE program aims for consistency and industry alignment in how to identify, define, and catalog publicly disclosed cybersecurity vulnerabilities in software. The CVE number can be universally used to track a particular vulnerability to a particular code base, and is referred to by vendors, researchers and consumers alike. Software projects affected by the vulnerability are able to contest this assignment, and there is scope for rejection of the CVE if warranted.

Often other types of metadata like CVSS (Common Vulnerability Scoring System) score and CWE (Common Weakness Enumeration) numbers are assigned during the security flaw triage process. They help in identifying the severity impact and applicability of the flaw, without necessarily studying the flaw in detail. They also help automated tools in processing flaws and assist security scanners.

Security scanners typically use CVE identifiers and CVSS scores to automatically identify and assess known vulnerabilities by matching detected issues against vulnerability databases. The CVSS scores provide standardized severity metrics, allowing scanners to prioritize flaws without detailed manual analysis. This enables efficient processing and remediation of security flaws, assists in determining their applicability to specific systems, and enhances reporting and compliance efforts through seamless integration with other security tools.

Given the dynamic nature of models and their inherent differences to traditional systems and application development, a "lift and shift" approach to security or safety management is not practical. However, there are valuable elements of existing security management processes that are well-tested, understood, and invested in across industry that could be applied to AI models allowing for appropriate security, safety, and risk management

## Scope of AI Security flaws

Security flaws, much like the skills and expertise required to manage them, exist differently than safety, bias, ethics or any other type of issues. In order to appropriately manage and resolve these for AI, we propose the following definition of an AI Security vulnerability.

**AI Security vulnerability:** *A flaw in an AI system, resulting from a weakness[22] that can be exploited, causing a negative impact on the confidentiality, integrity or availability of the affected component or components.*

Note: Here the word AI system is used instead of an AI model, because AI flaws tend to manifest themselves in the way the models are used. For example, in the case of a multi-modal model, it may generate unsafe images for some prompts, but if these models are never used by the application to generate images, but rather used for some other purpose, the flaw has not manifested into a vulnerability in that system.

While there are many similarities to security vulnerabilities, certain nuances only exist for AI models as a result of their nature and how they work, described below.

---

[22] A security weakness is a condition (i.e. flaw, error) or characteristic (i.e. improper configuration or weak encryption that permits the application or systems to have increased vulnerability to attack. See also https://nvd.nist.gov/vuln

- **Loss of Confidentiality:** When models respond with information they should not reveal as mentioned in its model card or other documents there is a loss of confidentiality. This includes but is not limited to:
    - Unauthorized PII and Intellectual property information.
    - Output that can cause widespread damage where the information is not generally available on the internet or considered public.
- **Loss of Integrity:** This issue can occur when an attacker can negatively influence the data within the model or is able to poison the model.
    - Adversarial fine-tuning can cause malicious changes to the model weights, and result in wrong or specially-crafted output generated by the model.
- **Loss of Availability:** The attacker is able to successfully perform a denial of service attack (DoS)[23] on the model[24] and stop all inference attempts. In most cases, this is considered a software issue rather than an AI model issue.

There are circumstances whereby the loss of confidentiality or integrity may result in human, social, or environmental harm, however this may only be ascertained in the context of the model's operational use. In such cases, the importance of information exchange between AI Security and AI Safety teams is paramount for fully exploring the impact of an exploited AI vulnerability so additional safeguards and security protections may be applied to protect against future occurrences of such incidents.

AI Security vulnerabilities do not take into account the platform on which the model is run, or the support libraries which are used to load, run and train the model. Security issues in those support infrastructures are already covered by the existing CVE process detailed above.

## Scope of AI Safety flaws

AI Safety is an interdisciplinary field focused on preventing accidents, misuse, or other harmful consequences arising from data generated from a generative AI model.

As discussed earlier, AI model safety is a novel concept as safety is not commonplace within the software development and security field, however there are a few noteworthy exceptions.  Medical Device Safety and Automotive Safety are well established within their industry with their own considerations for regulatory compliance beyond any risk management program. 'Trust and Safety' or 'Privacy and Trust' teams are often aligned with but executed and managed separately from Security resulting in their activities potential exclusion from an enterprise risk management program. These concepts have traditionally been excluded from enterprise risk management because their scope was limited to user privacy as a protected dataset due to regulatory requirements which today do not include the cognitive or mental impacts in exposure; though this is reflected in legal cases as "intangible injury" or non-economic damages for which software has no equivalent risk consideration or quantifiable measurement.

---

[23] https://arxiv.org/abs/2010.02432

[24] https://portswigger.net/daily-swig/deepsloth-researchers-find-denial-of-service-equivalent-against-machine-learning-systems

Technology producers that store user data, manage user information, or retain user account details often have to deal with adversarial attacks which target loss of confidentiality or loss of integrity. These attacks may include but are not limited to the compromise of user privacy (adversarial targeting of mail accounts of journalists), illicit account usage (storing and sharing explicit imagery of minors), spread of misinformation, and unauthorized sharing of sensitive content (false or true). Today, the technology industry's closest alignment with safety management is in regulatory duty to protect from harm, physically or cognitively (in the context of social platforms) and not necessarily in evaluating and quantifying the resulting impact from exposure of sensitive information, particularly from sources inappropriately perceived by the masses as trustworthy. Until recently, technology has been viewed as an enablement for harm such as that to occur but only because humans are the cause of violations in trust and safety (medical devices and automotive safety still remain the exception).

AI is however the first time in which the technology and its development are the cause of violations in trust and safety, bringing considerations of responsible and ethical AI to the forefront of the AI Safety discussion.

We propose to define AI Safety issues and flaws as *hazards* to avoid confusion between security flaw and safety flaw. We feel *hazard* more readily encapsulates the potential for harm separate from *flaw* or *issue* which are currently well utilized industry terms.

**AI Safety hazard:** An unexpected model behavior that is outside of the defined intent and scope of the model design. Safety hazards may result in harmful content generation, bias in decision making, or violation of social norms and ethics of groups; the impact and severity of which will vary greatly from group to group based on their culture, social, ethnic, or anthropic systems. Harm may be categorized by loss of life, injury or other physical or mental health impacts or damage, social and economic disruption or degradation, or some combination thereof.

## Taxonomy of AI Safety risks

An important part of handling AI Safety flaws is first to define and decide on a safety taxonomy which will be used by an AI model or an AI system. AI Safety taxonomy is a structured classification system that identifies, organizes, and informs different aspects of safety and reliability in artificial intelligence systems and their uses. The taxonomy helps outline various risks, challenges, and considerations that need to be addressed to prevent harm and ensure that AI systems operate as intended.

There is no shortage of AI Safety taxonomies and the research around it in recent years[25]. This paper does not aim to study all the taxonomies and does not intend to suggest which one should be used. However, some parameters for choosing a community taxonomy[26] would be:

- Taxonomy is available under a permissive open content license: This enables model makers and model distributors to use them to evaluate models, regardless of whether the model itself is open or closed.
- Taxonomy is developed in the open and anyone is free to contribute to its development: It does not end up being a research paper, but is in continuous development by the community.
- Taxonomy is extensible: You can use the taxonomy as the main document, and extend the taxonomy, adding variations as warranted based on the internal strategy of the model makers, provided the core elements of the taxonomy are met.
- A taxonomy does not cause harm. Taxonomies should be careful to not publish model responses to prompts because, for some hazard categories, these responses may contain content that could enable harm. Equally, unsafe responses could be used by technically sophisticated malicious actors to develop ways of bypassing and breaking the safety filters in existing models and applications. It may require some taxonomies to be society or culturally unique.[27]
- Taxonomies provide sufficient information to be used safely in benchmarking, in such a way that they are available for easy integration if required.

## Challenges

Tracking safety hazards has been one of the biggest challenges of AI models and systems so far. While there have been several developments in the safety field, such as safety aligned models that can detect unsafe content, a lack of consensus and understanding in the way safety hazards should be classified, tracked and perhaps even remediated persists which delays effective tracking, coordination, and mitigation of these hazards currently present in AI today.

---

[25] https://airisk.mit.edu/
[26] A community taxonomy is community-driven by the diverse members of the AI community helping to build and deliver safer and accurate AI systems.
[27] Taxonomies need to be socially or culturally aware. Certain references or expressions may be acceptable in one society or culture, but at the same time may be offensive to others.

## Handling AI Security and AI Safety

### Adapting vulnerability disclosure and coordination for addressing AI Security flaws

We are proposing the following adaptations for tracking AI related security flaws:

- Model providers should have a mechanism for reporting issues in a confidential manner.
- Reporters should disclose the security issue and all necessary details first and foremost to the model provider.
- Model providers have the primary responsibility to triage new reports and follow an established and public vulnerability disclosure process or policy. This is not without additional challenges[28].
- If the model provider is a CNA[29] and they recognize this as a security issue, then they alone are responsible for assigning a CVE ID to this issue. If the model provider is not a CNA, then the CVE id can be requested directly from the CVE program partner.[30]
- If the model provider does not recognise the issue as security related or if the model provider denies assigning a CVE, the reporter may raise a formal request with the CVE Program by following the instructions outlined on their site.[31]
- Model providers issue VEX statements to convey known vulnerability presence (or lack thereof) in models.

It should be noted that many of the current challenges that exist in the traditional vulnerability disclosure and coordination process have parallel challenges in AI Security. While models may publicly declare their scope and intent of use, triage must still be performed to ascertain the applicability, reproducibility, and impact of the reported flaw before coordination and remediation steps may be taken.

### Security frameworks and their relevance

To conclude this topic, we take a brief look at the various AI Security frameworks. These frameworks allow organizations to reduce the risk of the AI system and also help them meet regulatory requirements specific to the organization's industry vertical. It is usual for the industry to layer technical frameworks with those which are more conceptual.

Some of these frameworks are high level, some help to visualize the threat landscape, some of them help to meet the local regulatory requirements. In the end one would consider the proposed security handling procedure in this paper as a low level requirement/procedure,

---

[28] https://www.cve.org/Media/News/item/blog/2024/07/09/CVE-and-AIrelated-Vulnerabilities
[29] https://www.cve.org/ProgramOrganization/CNAs
[30] https://www.cve.org/PartnerInformation/ListofPartners
[31] https://www.cve.org/ReportRequest/ReportRequestForNonCNAs#RequestCVEID

which could very well be a subset of any of the above frameworks. We summarize these frameworks as follows:

Frameworks for Developers and Practitioners:
  – Focus: Practical security implementation, threat modeling, and vulnerability management.
  – Outcome: Secure AI models, prevent data breaches, and address risks in AI applications using tools like OWASP[32] and MITRE ATLAS.[33]

Frameworks for Chief Information Security Officers (CISOs):
  – Focus: Integration of AI Security into broader risk management.
  – Outcome: Align AI Security practices with existing cybersecurity programs and ensure compliance with regulatory standards (e.g., NIST AI RMF[34], Google SAIF[35]).

Legislature specific to a geographic area:
  – Examples: EU AI Act[36] and Canada's Artificial Intelligence and Data Act (AIDA[37]).
  – Outcome: Classify AI systems based on risk levels and provide governance rules to ensure safe, ethical use of AI, affecting compliance and legal responsibilities.

Current tracking efforts in the security community

There are several efforts ongoing in the security community in line with the above to streamline the process. Notable among them are the efforts by others for CVE/CWE for AI.[38]

## Proposal for hazards disclosure and management

Heavily drawing on the prior work done by Sven Cattell et al[39] This proposal is built up of two main components.

---

[32] https://owasp.org/www-project-ai-security-and-privacy-guide/
[33] https://atlas.mitre.org/
[34] https://www.nist.gov/itl/ai-risk-management-framework
[35] https://safety.google/cybersecurity-advancements/saif/
[36] https://www.europarl.europa.eu/topics/en/article/20230601STO93804/eu-ai-act-first-regulation-on-artificial-intelligence
[37] https://www.smartinsights.com/traffic-building-strategy/offer-and-message-development/aida-model/
[38] https://github.com/CWE-CAPEC/AI-Working-Group
[39] https://arxiv.org/pdf/2402.07039

Extending model/system cards

Model cards are used to document the possible use of the model, the architecture and sometimes the training data used to train the model. Today, they provide an initial set of human-generated material about the model that is beneficial in assessing its viability of use, however they have substantially more potential and applicability beyond their current usage. As organizations choose to adopt and build AI systems and capabilities built on top of models, they're choosing to do so in environments and with technologies they already have deployed. In order to facilitate better understanding of models wherever they travel or deploy to, we'd like to propose some changes to model cards.

The model card should serve as the introductory information about the model. In order for adopters and engineers to effectively compare models, we need a consistent set of minimum fields and content that must be present in a card. To provide for this, we suggest the development of a specification for model cards. Google's model card paper[40] was a novel first step in expressing this content in a standard way, and since 2018 industry has begun to ask different questions about models that warrant updates to this non-standardized format.

In addition to the existing fields recommended from the paper, we propose the following be added or modified:

- **Intent and Use:** While 'Intended Use' should describe the users (who) and use cases (what) of the model, it doesn't address how the model is to be used. Expanding Intended Use to be a statement for the model system that describes its usage with precise efficacy provides clarity in all three aspects of consideration (who, what, how) that is not present today. For example:
    - From a text prompt, we produce images that are safe for work, safe for children, and free of demographic bias
- **Scope:** The purpose of Scope is to exclude known issues that the Model producer has no intent or ability to resolve. While Intended Use should convey use cases that are out of scope, providing additional detail on specifics that are out of scope or consideration for resolution provides adopters or consumers of those models more context to make an informed decision from. This also ensures that reporters of hazards understand the purpose of the model before reporting a concern that is explicitly declared as unaddressable against its defined use. For example:
    - This is an LLM with no protections; prompt injections are out of scope.
- **Evaluation Data:** While an existing field in the model card, since 2018 we've developed evaluation frameworks[41] that focus on different considerations that inform adopters and consumers on what the model has achieved. We propose extending

---

[40] https://arxiv.org/pdf/1810.0399
[41] https://mlcommons.org/benchmarks/ai-safety/

Evaluation Data to provide a nested structure to convey if a framework was also used, and the outputs of such evaluation that were run on the model. Standardized safety evaluations are preferred and would enable a skilled individual to build a substantially equivalent model. Continued research and development to establish new or updated public evaluations should be encouraged.
- **Governance**: Governing information about the model is essential to understand how an adopter or consumer can engage with the model makers or understand the methodology by which it is produced..
- **References (optional):** Model manufacturers may find including references to be beneficial for potential model consumers in both understanding the model's operation, but also detailing artifacts and other content, such as an AIBOM, safety audit[42], or security audit, to demonstrate the maturity and professionalism of a given model.

Setting these and the existing fields as required elements for a model card allows for industry to begin to establish content that is essential for reasoning, decision-making, and reproducibility of models. By rendering model cards in an industry accepted format or standard, we promote the interoperability of models and their metadata across technology ecosystems.

We anticipate many organizations will build and develop applications and systems that embed, rely upon, and interact with models in an operational capacity. The loss of a model card as referenceable material for policy enforcement at runtime or auditing will exacerbate existing operational challenges in speedy, incident response and resolution. As we've seen with the supply chain security movement for software, the value of metadata and attestations about what has transpired to produce software is actionable for gatekeeping at deployment and in understanding where risk exists.

The data contained in model cards has the potential equivalent function, when stored in an widespread format, such as OC[43]I, to allow model cards, AIBOMs[44], and other evolving metadata types to be pushed, consumed, and leveraged in existing tooling and ecosystems with maturing processes for ingestion, analysis, insights, and policy enforcement.

---

[42] Safety audit is a comprehensive report identifying safety hazards, recommendations for remedial action and evidence of due diligence for stakeholders and regulators.
[43] https://opencontainers.org/
[44] https://owasp.org/www-project-aibom/

## Common flaws and exposures (CFEs) for Hazard tracking

While the Common Vulnerability disclosure mechanism used to track security flaws, is effective in traditional software security, its application to AI systems face several challenges:[45]

- AI models must satisfy statistical validity[46] thresholds[47] This means that any issues or problems identified in an AI model such as biases etc must be measured and evaluated against established statistical standards to ensure they are meaningful and significant.
- Concerns related to trustworthiness and bias often extend beyond the scope of security vulnerabilities and may not align with the accepted definition of security vulnerabilities.

Recognizing the above limitation we propose expanding the ecosystem, with a new term called "Common and Flaws and Exposure (CFE)", which is analogous to the CVE in the security space

### Coordinated Hazard Disclosure and Exposure (committee)

A central, neutral body must exist to track possible safety hazards in the same way security flaws are tracked (as discussed above). Reporters who discover safety issues, are expected to coordinate with the model providers to triage and do further analysis of these issues. Once established that the issue is indeed a safety hazard, this body assigns a CFE (Common flaws and exposure) number similar to CVE. Model makers and distributors can also request CFE numbers to track safety hazards they find in their models.

This body is the custodian of CFE numbers. They are responsible for assigning them to safety hazards as per the process described below, for tracking them and if at some point publishing them in various forms. We propose a "HEX" format to publish CFE data in the section below.

### Adjunct panel

In accordance with the paper by Sven Cattell et al, we would also like to propose an adjunct panel to facilitate resolution of safety hazards that are contested:

---

[45] These details are taken directly from the paper: https://arxiv.org/pdf/2402.07039
[46] Refers to the confidence that the results or conclusions drawn from the model's performance are not due to random chance but are supported by robust statistical evidence.
[47] Criteria or benchmark

If a reporter submits a safety hazard on statistical or other grounds, a vendor or model maker may reject or contest the safety hazard and the Adjunct Panel is informed. In the course of reporting, if the submitter-supplied sample/output pairs are statistically biased, and an unbiased sample set would not show a violation of the model card, the adjudicator panel may request supporting data from the vendor or model maker to validate the rejection. The adjudicator and model maker can jointly assess whether the maker-supplied data is too sensitive for the submitter and determine on further steps for resolution.

In the initial phase of implementation, bootstrapping the Adjunct panel with the Coordinated Hazard disclosure committee (described above) provides the initial expertise and alignment for future engagement between the two bodies. However it is important for each body to be neutral and/or have good representation from major vendors/reporters in the ecosystem.

### Adapting VEX

The Vulnerability Exploit eXchange format (VEX)[48] is a recent format for conveying the exploitability status of a vulnerability in software. While nascent in industry adoption, we believe this style of information conveyance can be leveraged in the coordinated disclosure and necessary resolution of AI Safety hazards.

We propose exploration into what a potential Hazards Exposure eXchange (HEX) format would provide the industry in conveying the exposure and resulting impact a safety hazard has on the operational use of AI. The impact of those hazards on intended model use and outputs must be tracked to inform consumers how their use is impacted so they may take steps to correct any decisions or harm resulting from an impacted AI system, this includes hazards resulting from the development, training, and inference of the model and all its variants.

Most existing VEX fields and values are adaptable to safety hazards.. Where VEX defines the subcomponent ID, we propose an unique identifier construct (such as a commit) and category (such as the model lifecycle stage and source) to define how and where the hazard is introduced. Status of the hazard is a critical field in providing consumers of models with sufficient information to understand how they have been exposed and must consider content from the model card (intended use) in detailing that status (i.e. affected, unaffected, fixed, under investigation) and justification (i.e. model_use_not_approved, guardrails_in_place, or tuned_out). Some models provide for multiple uses in which a hazard may only adversely impact a portion of them, it is important to inform model consumers if their use matches an impacted use and what the specifics of that impact are just as much as understanding how they may not be affected by the hazard in their use of the model.

---

[48] https://www.cisa.gov/resources-tools/resources/minimum-requirements-vulnerability-exploitability-exchange-vex

Additional research is required to develop status justification statements and other potential HEX fields that are informed by CFE and Safety Frameworks in order to comprehensively develop a proposal for HEX acceptable to industry.

## Workflow

We would like to propose a workflow similar to the security workflow above, with several marked differences to address considerations highlighted previously regarding AI Safety

- Reporters attempting to report safety issues would do so in accordance with the specified intent and scope in the model card, reflective of safety benchmarking that has occurred. If, for example, the model is not intended for a specific use, has had a topic explicitly excluded from training or is considered beyond the scope of the model, the safety issue being reported is likely to be closed as invalid by the model maker.
- Safety issues are reported to the model makers, following their established and discoverable safety reporting process. If they acknowledge they can approach the Coordinated Hazard Disclosure committee for a CFE number, which is then used to track that particular safety issue.
- If model makers refuse to acknowledge the safety issue, reporters can approach the Adjudicator panel, which is responsible for resolving disputes.
- In the end, the issue is either accepted or rejected. If accepted, the issue is published in a public database and the model maker issues a safety advisory pointing to the public record.

As the process evolves, several enhancements may be made to extend the safety tracking and root causes of the safety hazards to build industry knowledge of prevention, akin to data like CPEs (Common Platform Enumeration) or CWEs.

### Current safety efforts

At this point authors would like to mention several community efforts currently underway for standardization of processes. Some of them may be orthogonal to each other, but we would like to stress the need to collaborate here, rather than compete. The list below is in no way exhaustive, and it's quite possible that new efforts would rise or the ones listed here would become inactive after this paper is published:
- AI Alliance – Trust and Safety group[49]

---

[49] https://thealliance.ai/

- MLCommons AI Safety group[50]
- Coalition of Secure AI[51]

## Current challenges in the industry for implementation of this proposal

Several challenges in the current model ecosystem exist and they need to be addressed in order to implement open and effective tracking of safety hazards.

- **Model/system cards:** While most bigger models have explicit model cards, containing details of how the model has been trained and output of the various benchmarks, for other models, the process is not uniform across the industry and has to improve. Furthermore the model cards need to evolve to include both the intent as well as the scope of the model, in order to ensure that reporters understand the purpose of the model before reporting safety flaws.

- **Clear distinction between security and safety issues:** Currently there is no distinction between security and safety issues for most reporters or researchers. When reporters and researchers discover safety hazards or security vulnerabilities with an AI system, they'll look for a central reporting mechanism to submit the report to. Since these suspected issues may not be immediately distinguishable as security or safety impacting, the triage and impact assignment (security versus safety) may fall to the model-maker or producer to ascertain, unless the reporter can distinctly identify the source of the issue, such as with safety hazards in scope of the model card. Safety and Security reporting mechanisms may need to combine necessary reporting fields to enable triaging to occur for assignment and validation by an AI Safety or AI Security team, as appropriate.

- **Model maker recognition of safety hazards:** Most model makers clearly advertise what issues they would consider to be in the scope for security flaws[52], there is often little or no discussion about safety issues and how the model makers intend to treat safety reports around them. Setting clear expectations about handling safety issues would help both the model users and reporters.

- **Standardization of safety evaluations:** To facilitate easier and more meaningful comparisons between AI models, we propose establishing a standardized specification for safety evaluations. Currently, model makers use various benchmarks and safety leaderboards to evaluate their models, which makes it challenging to compare one model with another due to the lack of uniform criteria. Moreover, these leaderboards often lack independent verifiability and reproducibility against established frameworks or benchmarks.

---

[50] https://mlcommons.org/working-groups/airr/ai-risk-and-reliability/
[51] https://www.coalitionforsecureai.org/
[52] https://www.anthropic.com/responsible-disclosure-policy

By defining the minimum required fields and criteria that a safety evaluation must cover, we can enable consistent, transparent, and reproducible safety assessments across different AI models. This standardization would help in independently verifying results and ensure that safety evaluations are comprehensive and comparable.

## Conclusion

Models released publicly to the community and developed according to open source principles could play a significant role in the future of AI models. Frameworks and tools necessary for developing and managing models against industry and consumer expectations, requires openness for consistency and to provide equitable content of consideration in making organizational risk decisions. By increasing the transparency and access to elements that are critical to the functionality and output content of the models (such as the source data), the greater industry's ability to discover, identify, track, and resolve hazards and vulnerabilities before they have widespread impact. As we've seen with the widespread inclusion of open source in software, we can reasonably expect the future of AI models to follow a similar path. While the majority of current efforts made by model makers are concentrated on security features, safety is an equally important aspect for responsible, ethical and trustworthy use of these models. The proposals presented in this paper are intended to afford flexibility through consistent means of governance, workflows, and structure that have been in use within the security community for years. When implemented, they may provide a shortcut without compromise to resolve the pressing need of managing AI Safety effectively and rapidly.

Lastly, while several disjointed efforts exist currently, there is a need for collaboration between models producers, consumers, legislative bodies, and law enforcement agencies to streamline this process.

https://cdn.openai.com/research-covers/language-unsupervised/language_understanding_paper.pdf

Deep Unsupervised Learning using Nonequilibrium Thermodynamics, Jascha Sohl-Dickstein, Eric A. Weiss, Niru Maheswaranathan, Surya Ganguli. https://arxiv.org/abs/1503.03585, 2015

Concrete problems in AI Safety, Amodei, D., Olah, C., Steinhardt, J., Christiano, P., Schulman, J., & Mané, D. (2016).. arXiv preprint arXiv:1606.06565.

# Definitions

**AI Ethics**: A field of AI that is focused on evaluating the fairness, bias, governance, and responsible creation, distribution, and use of AI. It is considered a subset of AI Trustworthiness.

**AI model:** An AI model is a software framework that learns from data to make predictions or decisions, simulating aspects of human cognitive processes. *Source: GENAI Commons[53]*

**AI Security**: The field of security focused on protecting and securing AI, AI Systems, and AI workloads. It covers the security of: data and the model was trained on, the supply chain of the AI Model, security capabilities that support the operational security of models such as prompt injection protection and data exfiltration detection, and other related areas.

**AI Safety**: The field of study and practices to ensure AI systems operate in a manner that is safe and aligned with human values, preventing harm to individuals and society. It is considered a subset of AI Trustworthiness. *Source: Modified for Red Hat from GENAI Commons*

**AI system**: LLMs and everything support libraries and software, for example pytorch + inference software. RHEL AI is considered an AI system, as it is packaged to include the necessary components and software for use.

**AI Trustworthiness**: The collection of characteristics, measurements, and corresponding verifications regarding the explainability, reliability, and safety of AI and its use in real-world applications. It considers the design and operation of AI to ensure it is ethically sound, legally compliant, and reliable. Typically encompassing principles of transparency, fairness, and accountability. AI Trustworthiness is focused on the "what" - generating sufficient information to establish trust in a model against expectations. *Source: Modified for Red Hat from GENAI Commons*

**Embargoed security flaws:** Embargoed security flaws are vulnerabilities or weaknesses in software, hardware, or systems that have been identified but are not yet publicly disclosed. Information about these flaws is temporarily restricted to a limited group of trusted parties—such as developers, security teams, and affected vendors—under an agreement not to share details externally until a specified date or event. The purpose of the embargo is to allow time for patches or fixes to be developed, tested, and distributed before the vulnerability becomes widely known.

---
[53] https://genaicommons.org/glossary/ai-model/

**LLM Guardrails:** Guardrails are the set of security and safety controls that monitor and dictate a user's interaction with a LLM application. They are a set of programmable, rule-based systems that sit in between users and foundational models in order to make sure the AI model is operating between defined principles in an organization.

**Model Cards:** A model card is a type of documentation that is created for, and provided with, machine learning models. A model card functions as a type of data sheet, similar in principle to the consumer safety labels, food nutritional labels, a material safety data sheet or product spec sheets.

**Responsible AI:** the development and use of AI in a way that is ethical, transparent, and accountable. Responsible AI involves ensuring that AI systems are designed with consideration for their potential impact on individuals and society. It emphasizes the importance of creating AI systems that are non-discriminatory, fair, privacy-respecting, and secure. It also involves adhering to regulations and guidelines, engaging in ethical decision-making processes, and ensuring that AI technologies benefit humanity as a whole. Responsible AI is focused on the "how" of AI - how it was developed and used against egalitarian principles for human benefit, whereas the AI trustworthiness is focused more on the "what". Source: Modified for Red Hat from GENAI Commons

**Security Weakness:** A weakness is a condition in a software, firmware, hardware, or service component that, under the right circumstances, could contribute to the introduction of vulnerabilities.

**Safety Hazard**: An unexpected behavior or output outside of the defined intent and scope of a system's or software's design. Safety hazards are linked with producing harmful content or outputs that can cause social, economic, and environmental harm to consumers and users.

**Sustainable AI**: the development and use of carbon neutral and carbon negative practices to minimize the negative environmental impacts of AI technologies. While considered its own independent domain, the ecological and corresponding human harm that results from unsustainable AI development and use aligns with Responsible AI – directly influencing the duration of benefits for humans against environmental realities (akin to corporate social responsibility focuses). It is related to AI Ethics.